\documentclass[aps,prb,twocolumn,groupedaddress,showpacs,superscriptaddress,amssymb,amsmath]{revtex4-1}
\usepackage{graphicx}
\usepackage{tabularx}
\usepackage{color}
\usepackage{amsmath}
\usepackage{tikz}
\usetikzlibrary{tikzmark,fit}
\usepackage{comment}
\newcommand{\bea}{\begin{eqnarray}}
\newcommand{\eea}{\end{eqnarray}} 

\newcommand{\be}{\begin{equation}} 
\newcommand{\ee}{\end{equation}}
\usepackage{dcolumn}
\usepackage{hyperref}
\usepackage{bm}
\usepackage{epsf}
\usepackage{subfigure}
\usepackage{epstopdf}%
\usepackage{amsfonts}

\begin{document}

\title{Harmonic chains and the thermal diode effect}

\author{Na'im Kalantar}
\affiliation{Department of Chemistry,
University of Toronto, 80 Saint George St., Toronto, Ontario, Canada M5S 3H6}

\author{Bijay Kumar Agarwalla}
\affiliation{Department of Physics, Dr. Homi Bhabha Road, Indian Institute of Science Education and Research, Pune, India 411008}

\author{Dvira Segal}
\email{dvira.segal@utoronto.ca}
\affiliation{Department of Chemistry and Centre for Quantum Information and Quantum Control,
University of Toronto, 80 Saint George St., Toronto, Ontario, Canada M5S 3H6}
\affiliation{Department of Physics, University of Toronto, Toronto, Ontario, Canada M5S 1A7}

\date{\today}
\begin{abstract}
Harmonic oscillator chains connecting two harmonic reservoirs at different constant temperatures cannot act
as thermal diodes, irrespective of structural asymmetry. 
However, here we prove that perfectly harmonic junctions can rectify heat 
once the reservoirs (described by white Langevin noise) 
are placed under temperature gradients, which are asymmetric at the two sides, an effect that we term 
``temperature-gradient harmonic oscillator diodes".
This nonlinear diode effect results from the additional constraint---the imposed thermal gradient 
at the boundaries. 
We demonstrate the rectification behavior based on the exact analytical formulation of steady state heat transport in harmonic systems coupled to Langevin baths, 
which can describe quantum and classical transport, 
both regimes realizing the diode effect under the involved boundary conditions.
Our study shows that asymmetric harmonic systems, such as room-temperature hydrocarbon molecules 
with varying side groups and end groups, or a linear lattice of trapped ions may 
rectify heat by going beyond simple boundary conditions.
\end{abstract}

\maketitle 

\section{Introduction}
\label{Sec-intro}

Energy transport processes play central roles in chemical reactivity, 
biological function, and the operation of mechanical, electronic, thermal, and thermoelectric devices 
\cite{RubtsovRev,BijayRev,ReddyRev}. 
Understanding energy transport in both the classical and quantum regimes is fundamental to
thermodynamics, relaxation dynamics, chemical reactivity, and biomolecular dynamics 
\cite{DharRev,BijayRev,LeitnerRev}. 

Linear, one dimensional (1D) chains of particles and springs serve to model vibrational (phononic) heat transport 
through molecular chains. 
The force field, the functional form of the of potential energy and its parametrization is 
often constructed by hand, such as in the eminent FPU model, 
to represent basic harmonic and anharmonic interactions \cite{FPU1,FPU2}. 
In molecular simulations, the force field is taken from first-principle (DFT) calculations \cite{Pauly}.
Recent experiments probed the flow of vibrational energy (heat) through self-assembled monolayers 
of alkanes \cite{Braun,Gotsmann,Malen} down to a single molecular junction \cite{Brendt,ReddyE}. 
These junctions comprise a linear (quasi 1D) molecule bridging two solids with the steady state thermal 
heat current or the thermal conductance as observables of interest. 

When the temperature is low relative to the characteristic vibrational frequencies, 
the harmonic force field can be adopted to model interactions
in molecules since atomic displacements stay close to equilibrium.
However, the harmonic potential leads to several intriguing, anomalous properties:
Heat current in harmonic chains was calculated in both the classical 
\cite{Lebowitz} and quantum \cite{Rego98,Segal03} regimes displaying 
an anomalous thermal conductivity that was diverging with size, 
in disagreement with the phenomenological-macroscopic 
Fourier's law of heat conduction \cite{DharRev}. 

Purely harmonic systems connecting harmonic baths at fixed temperatures $T_H$ and $T_C$
cannot support the thermal diode effect: The heat current is exactly
symmetric upon exchange of temperatures between the heat source and drain, 
as directly observed from the Landauer formula for heat conduction 
\cite{Rego98,Segal03,DharRev}. 
Recent studies realized a diode effect in harmonic junctions---by making parameters
to be temperature-dependent---thus sensitive to the direction of the thermal bias
\cite{commentMuga}.
Fundamentally, such effective harmonic models emerge due to underlying nonlinear interactions.

The thermal diode (rectifier) effect had been demonstrated in numerous 1D chains by combining
anharmonic interactions and spatial asymmetry starting from Refs. \cite{casati02,Li04}. 
In one type of modelling, the chain is made of different segments and the diode effect can be explained due to the mismatch 
in the phonon spectral density in the forward and backward temperature-bias directions. 
Thermal rectifiers were further proposed in other models based on classical 
\cite{BaowenRev,BenentiR,WongRev} 
and quantum transport equations \cite{SB,Bijay17, BenentiR,PereiraRev, Leitner13,Leitner19},
with recent efforts dedicated to achieving high rectification ratios that persist with length
\cite{CasatiEPL,Chen,Alexander,You}.

At the nanoscale, the key ingredients of a thermal diode realized with
harmonic reservoirs are (i) structural asymmetry, 
e.g. by using graded materials and (ii) anharmonicity of the force field 
\cite{Wu09,Claire09,LeitnerRev}.  
Anharmonicity in the form of a two-state system  \cite{SB,Bijay17} can be readily
realized in hybrid models with an impurity \cite{Claire09} or
 spin chains coupled to boson baths \cite{XXP,Vinitha}
(as well as in the opposite scenario of a boson chain coupled to spin baths \cite{Goold});
recent experiments demonstrated heat rectification with
 Josephson junction qubits \cite{Pekola,Haack}.

In contrast, in molecules such as alkane chains the harmonic force field dominates 
interactions at room temperature. Therefore, these molecules do not realize a noticeable 
diode effect in a steady state solid-molecule-solid configuration when constant
temperatures  ($T_H$ and $T_C$) are  maintained at the boundaries  \cite{Segal03}.
Pump-probe transient spectroscopy experiments demonstrated unidirectional 
vibrational energy flow between different chemical groups
(e.g., nitro and phenyl) \cite{Dlott1,Dlott2};  corresponding observations of steady state asymmetric heat flow through molecules 
are still missing \cite{CNTE}.

\begin{figure} 
\includegraphics[scale=0.38]{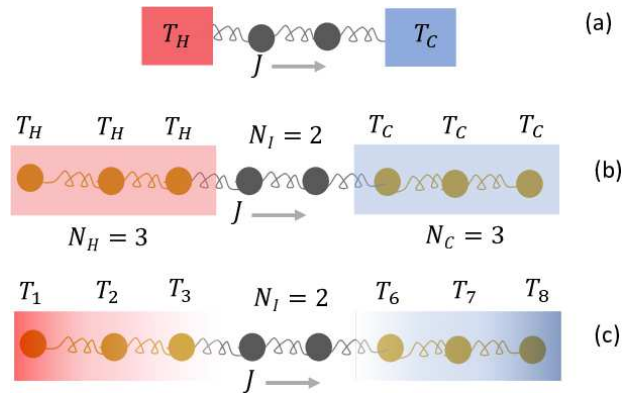}
\caption{
(a) $N$-particle chain connecting two heat baths (modeled by Langevin thermostats), 
hot and cold, with bead 1 connected to the hot bath and bead $N$ coupled to the cold one.
In this example, $N=2$.
(b)-(c) $N$-site chain made of $N_H+N_C$ exterior beads coupled to Langevin heat baths
and $N_I$ particles in the central, interior zone.
The imposed temperature at the edges may be homogeneous as in (b), or inhomogeneous as in (c), 
the latter potentially realizing a TGHO diode. 
In (b)-(c), we used $N=8$ total number of beads.
}
\label{Fig1}
\end{figure}

Can harmonic systems support the diode effect? 
In this paper, our goal is to revisit the problem of steady state heat transfer in asymmetric harmonic 
junctions and make clear the conditions for the realization of a thermal diode effect.
In our model all components are harmonic: the reservoirs, representing e.g. solids,
the chain (molecule), and their couplings. Furthermore, we do not 
effectively include anharmonicity by making parameters temperature dependent.
As we had just discussed, microscopic harmonic chains that bridge two harmonic solids, a heat source and a heat drain at constant temperatures $T_H$ and $T_C$, respectively, as depicted in 
\ref{Fig1}(a)-(b), cannot act like a diode irrespective of structural asymmetry. 
However, once we modify the boundary condition as we show in  Fig. \ref{Fig1}(c)
and impose {\it thermal gradients} in the 
contact region, 
the  junction can {\it rectify} heat due to the (multi-affinity) boundary conditions, 
with particles directly coupled to different baths. 

We exemplify this scenario, referred to as the temperature-gradient harmonic oscillator (TGHO) chain in 
Fig. \ref{Fig1}(c). The hot solid is divided into several regions with externally controlled temperatures, $T_1^H>T_2^H>T_3^H$. Similarly, 
the colder region may be divided into domains with externally-controlled temperatures.
This setup can be realized experimentally by controlling local temperatures 
(as in trapped-ions chain in optical lattices \cite{Haffner}), or computationally, 
as a mean to introduce thermal gradients in structures, the result of genuine inelastic scatterings. 

Our analysis is performed using formally-exact expressions for the heat current based on 
the quantum Langevin equation \cite{DharRev}. 
Both classical and quantum harmonic diodes are demonstrated, 
with quantum effects leading to an improved performance of the TGHO diodes. 
Furthermore, we describe a unique, purely-quantum TGHO diode, which does not have a classical analogue.
As for classical diodes, 
we perform classical molecular dynamics (MD) simulations 
of heat flow in anharmonic junctions to demonstrate the extent of the diode effect under explicit anharmonicity in comparison to the TGHO diode.

Altogether, in this work we:
(i) Derive conditions for realizing 
a new type of thermal diodes, the TGHO diode
based on structural asymmetry and inhomogeneous temperature boundary conditions, 
(ii) identify a purely-quantum TGHO diode,
(iii) make clear  conditions for realizing thermal 
diodes in either genuine or effective harmonic models.

%


\section{Model and Method: Linear chain coupled to heat baths}
\label{Sec-M}

\subsection{Model}

We focus on a 1D harmonic oscillator chain with a total of $N$ beads.
The chain is coupled at its to edges to two thermostats, also referred to as solids.
In simulations of heat transport through 
solid-molecule-solid junctions, typically,
rather than including the solids' atoms explicitly, they are emulated through Langevin baths 
to which the first and last atoms of the molecule are attached, see Fig.~\ref{Fig1}(a).
This setup was considered in numerous computational studies, see e.g. \cite{Segal03,SB,DharRev}, 
and since the system as a whole is microscopically harmonic, it cannot support the diode effect.

Let us now consider a more complex picture of a junction with {\it several} 
beads on each side ($N_H$, $N_C$) each
attached to an independent Langevin noise term. 
The $N_I$ interior particles are not thermostated.
For example, in Fig.~\ref{Fig1}(b)-(c) we display an $N=8$-bead chain where atoms 1, 2 and 3 coupled 
to hot baths, while beads 6,7, and 8 connected to colder reservoirs. 
We can think about this scenario in two different ways:
We may regard all $N$ beads as part of the molecular system, with the heater and sink reservoirs 
(implemented via Langevin noise) acting on several edge sites.
Alternatively, we can picture this setup as a molecule made of the $N_I$ interior beads only 
(4 and 5), with the modelling of the thermal reservoirs enriched: 
The solids are described by $N_H$ and $N_C$ physical beads, 
each connected to an independent Langevin bath. In fact, this latter approach has been 
adopted in molecular dynamics simulations of thermal conductance of nanoscale systems. 
It allows to engineer a nontrivial phonon spectral function within a 
standard (white noise) Langevin simulation method \cite{Inon1}.

What about the temperatures imposed at the boundaries? We consider two cases:
(i) The temperature is homogeneous at the edges, $T_H=T_{1,2,3}$ and $T_C=T_{6,7,8}$.
That is, beads 1, 2, and 3 are coupled to three independent Langevin baths, but each is maintained at the same temperature (and similarly for the cold side). 
This scenario is depicted in Fig.~\ref{Fig1}(b).  
(ii) A  temperature profile is implemented at the edges: 
beads 1, 2, and 3 are coupled to Langevin baths with a {\it thermal gradient} 
such that the temperature of the attached baths follow the trend 
$T_1>T_2>T_3>T_4>T_5>T_6$, see Fig.~\ref{Fig1}(c). 
It is not required that all temperatures vary; 
at minimum we require two affinities (three baths of different temperatures). 
We refer to this scenario as the temperature gradient harmonic-oscillator chain.

In what follows, we show that these two cases 
are {\it fundamentally} distinct. In the first setup, Fig.~\ref{Fig1}(b),  a diode effect 
{\it cannot} show up even under structural asymmetries; 
remember that we work with harmonic oscillators.
In contrast, in the second scenario, Fig.~\ref{Fig1}(c), a diode effect develops 
in both the classical and quantum regimes when the gradients are distinct
 and structural asymmetry is introduced. 
Moreover, we show that in a certain setup, a TGHO chain can 
support a purely-quantum diode---with no corresponding classical analogue.

\subsection{Langevin equation formalism}

We write down the classical Hamiltonian and corresponding classical equations of motion (EOM); a quantum description based on Heisenberg EOM directly follows  \cite{DharRev},
\bea
H= \sum_{i=1}^N\frac{p_i^2}{2m_i}  + \frac{1}{2}\sum_{i=1}^{N+1}k_{i-1}(x_i-x_{i-1}-a)^2.
%
\label{eq:H}
\eea
Here, $x_0$ and $x_{N+1}$ are fixed, setting the boundaries. 
$a$ is the equilibrium distance between nearest-neighbor sites.
%

At this stage, we assume that every particle $i$ is coupled to an 
independent heat bath. This coupling
is incorporated using the Langevin equation with a friction constant $\gamma_i$ 
and stochastic forces $\xi_i(t)$ obeying the fluctuation-dissipation relation associated 
with exchanging energy with a heat bath,
$\langle \xi_i(t) \xi_{i'}(t')\rangle=2T_i\gamma_i \delta(t-t')\delta_{i,i'}$.
In the model for the diode below, we specify the interior region (which is not thermostated)
by setting its friction constants to zero. However, the TGHO effect is generic 
and can be discussed even when every bead is attached to a thermostat. 

The classical EOM for the displacements are 
\bea
m_i\ddot {x}_i&=& -k_{i-1}(x_i-x_{i-1}-a) + k_{i}(x_{i+1}-x_i-a)  
\nonumber\\
&-&\gamma_i v_i + \xi_i(t),
\eea
with $v_i$ as the velocity of the $i$th particle.

The steady state heat current can be evaluated inside the chain by calculating
heat exchange between beads, or at the contact region with each bath. Using the latter approach,
the classical (C) heat current from bath $l$ to its attached bead is ($k_B\equiv 1$, $\hbar\equiv1$), \cite{DharRev}
\bea
J_l^C=\sum_m\gamma_l\gamma_m \int_{-\infty}^{\infty} d\omega \frac{\omega^2}{\pi} |(G(\omega))_{l,m}|^2(T_l-T_m).
\label{eq:JC}
\eea
The summation is done over every thermostat. In what follows, we introduce the compact notation 
\bea
M_{lm} \equiv \gamma_l\gamma_m\int_{-\infty}^{\infty} d\omega \frac{\omega^2}{\pi} |(G(\omega))_{l,m}|^2,
\label{eq:M}
\eea
and write down $J_l^C=\sum_m M_{lm}(T_l-T_m)$.

It can be shown that Eq. (\ref{eq:JC}) generalizes in the quantum (Q) case
to \cite{DharRev}
\bea
J_l^Q=\sum_m\gamma_l\gamma_m \int_{-\infty}^{\infty} d\omega \frac{\omega^3}{\pi} |(G(\omega))_{l,m}|^2[n_l(\omega)-n_m(\omega)],
\nonumber\\
\label{eq:JQ}
\eea
with $n_l(\omega)=[e^{\omega/{T_l}}-1]^{-1}$, the Bose-Einstein distribution function of bath $l$ of temperature $T_l$.
Here,  $\boldsymbol{G}(\omega)$ is a symmetric matrix. 
%
The matrix $\boldsymbol{G^{-1}}(\omega)$ for the five-site model that we simulate below
is given in Appendix A.  

To calculate the net heat current, we separate the heat baths into two groups, 
$N_H$ heat sources placed to the left of the interior region,  and $N_C$ heat sinks at the other side.
The total input heat power is 
\bea
J=\sum_{l=1}^{N_H}{J_l},
\label{eq:Jtotal}
\eea
and it equals the total output heat current at the colder baths.

We now reiterate that a thermal diode effect cannot appear in 
harmonic chains coupled to heat baths at two different temperatures (single affinity setup). 
If $N_H$ beads are coupled to heat baths at $T_H$ and similarly, 
$N_C$ beads are attached to reservoirs at temperature $T_C$,
the net quantum heat current is given by
%
$J^Q=\sum_{l \in{N_H}} \sum_{m \in {N_C}}\gamma_l\gamma_m \int d\omega \frac{\omega^3}{\pi} |(G(\omega))_{l,m}|^2[n_H(\omega)-n_C(\omega)]$.
%
This expression is symmetric under the exchange of temperatures
even if long range interactions are included so that $\boldsymbol{G}(\omega)$ is a full matrix. 
Thus, this setup cannot support a diode effect.
The multi-affinity scenario 
is discussed in the next section.

\vspace{5mm}
\begin{figure} 
\includegraphics[width=.45\textwidth]{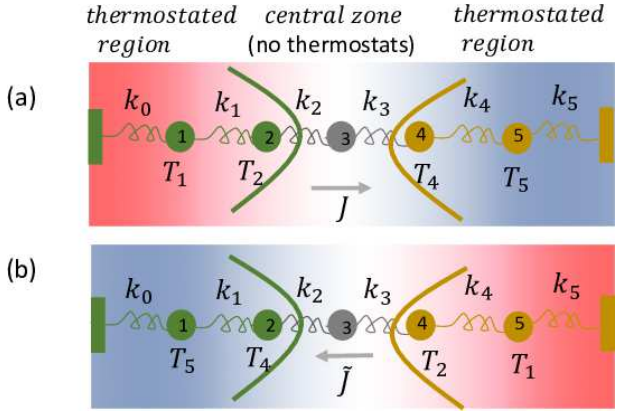} 
\caption{
A thermal rectifier based on an $N=5$-bead harmonic chain. 
Two beads at the boundaries are considered part of the solids, and they
directly exchange energy with Langevin thermostats.
(a) In the forward direction we set $T_1>T_2>T_4>T_5$ and calculate the total heat input $J$ from the baths attached to sites 1 and 2.
(b) In the backward direction we interchange the temperatures such that bead 1 (2) is now attached to a thermal bath at temperature $T_5$ ($T_4$), and similarly for the other half. 
In this case we calculate the total heat input $\tilde J$ from the hot baths, attached now to beads 4 and 5.
}
\label{Fig2}
\end{figure}

\section{TGHO diodes}

In this Section, we describe the principles behind the TGHO diode. 
We begin by exemplifying this effect in an $N=5$-bead chain depicted in Fig.~\ref{Fig2}, 
then we generalize the discussion to longer systems.  
As a case study, we set $N_I=1$, $N_H=N_C=2$; beads 1 and 2 are connected to hot baths, 
beads 4 and 5 are coupled to colder reservoirs, the central bead 3 is not thermostated. 
This separation is arbitrary and in practice should be based on the physical structure. 

We begin with the classical (C) limit, Eq. (\ref{eq:JC}). 
The total heat input in the forward ($J$) direction, corresponding to the setup of Fig.~\ref{Fig2}(a) is
\bea J^C = (T_1-T_4)M_{14} + (T_2-T_5)M_{25} \\ \nonumber + (T_1-T_5)M_{15} + (T_2-T_4)M_{24}. 
\eea 
Reversing the temperature profile as in Fig.~\ref{Fig2}(b), $T_{1}\leftrightarrow T_5$ and $T_2\leftrightarrow T_4$, the reversed ($\tilde J$) current is
\bea 
\tilde J^C = (T_5-T_2)M_{14} + (T_4-T_1)M_{25}
 \\\nonumber + (T_5-T_1)M_{15} + (T_4-T_2)M_{24}. 
 \eea
%
The sum of the opposite currents, which quantifies the diode effect is
\bea\Delta J &\equiv& J^C + \tilde J^C 
\nonumber\\
&=& \left[(T_1 - T_2) - (T_4 - T_5)\right](M_{14} - M_{25}).
\label{eq:DeltaJ}
\eea
We can now identify the necessary conditions for realizing the diode effect, $\Delta J \neq 0$: 
(i) The temperature gradients should be {\it distinct at the two boundaries}, 
$(T_1-T_2)  \neq (T_4 -T_5)$.
(ii) The setup should include a spatial asymmetry such that 
$M_{14} \neq M_{25}$. Asymmetry should be introduced in the thermostated region, as we prove next.
Explicitly, assuming the friction constants are uniform, $\gamma_{1,2,4,5}=\gamma$, we get (Appendix A):
 \bea M_{14} &=&          
 \frac{\gamma^2}{\pi}  \int_{-\infty}^{\infty} d\omega \omega^2\frac{\lvert k_1k_2k_3\left(-\omega^2 + i\gamma\omega + k_4 + k_5\right)\rvert^2}{\lvert\det \boldsymbol{G^{-1}}\rvert^2}
 \nonumber\\
 M_{25} &=&
 \frac{\gamma^2}{\pi}  \int_{-\infty}^{\infty}  d\omega \omega^2\frac{\lvert k_2k_3k_4(-\omega^2+i\gamma\omega + k_0 + k_1)\rvert^2}{\lvert\det
 \boldsymbol{ G^{-1}}\rvert^2}.
 \nonumber\\
 \label{eq:M14M25}
 \eea
Therefore, asymmetry in the central zone (see definitions in Fig.~\ref{Fig2}), 
in the form $k_2\neq k_3$ cannot lead to the required asymmetry $M_{14}\neq M_{25}$, since these terms are not sensitive to the asymmetry.
For the diode effect to hold, structural asymmetry must be included in the {\it thermostated zones}.
For example, it could be introduced in the form  $k_{1}=k_0\neq k_4=k_5$.
In appendix A we consider chains of arbitrary size $N_I$, with $N_H=N_C=2$ and
prove that structural asymmetry must be introduced within the
thermostated zones to realize a diode.

Furthermore, in a chain of length $N$ with $N_B$ beads in each thermostated zone, 
\bea 
J^C &=& \sum_{i=1}^{N_B} \sum_{j=1}^{N_B}(T_i - T_{N+1-j}) M_{i,N+1-j}  
\nonumber\\
\tilde J^C &= &\sum_{i=1}^{N_B}\sum_{j=1}^{N_B} (T_{N+1-i}-T_j) M_{i,N+1-j}.
\eea
Therefore,
\bea
\Delta J = \sum_{i=1}^{N_B}\sum_{j\neq i}\left[(T_i - T_j) + (T_{N+1-i} - T_{N+1-j})\right] M_{i,N+1-j}
\nonumber\\
\eea
Physically, the two asymmetries (structural and in the applied thermal gradients) 
are achievable in molecular junctions by connecting a molecule to distinct solids:
Different materials are characterized by different phonon properties
such that the force constants at the left side would be distinct from those at the right side,
leading to the required spatial asymmetry (ii).
Furthermore, given that different materials are employed at the two sides, 
it is reasonable to assume that a total imposed gradient $\Delta T$ 
would be divided unevenly on the two boundary regions such that condition (i) is satisfied. 
(In real materials, these gradients develop due to lattice anharmonicity.)  
Most importantly, we reiterate that 
imposing structural asymmetry ($k_2\neq k_3$ in Fig. \ref{Fig2}) 
while using identical boundaries ($k_0=k_1=k_4=k_5$) cannot result in thermal rectification in our
model. 

In Appendix B, we discuss the corresponding TGHO diode effect for harmonic chains
with local trapping (pinning) potentials. We show that the TGHO diode effect can develop only 
once pinning potentials at the two thermostated regions are different--applying 
as well unequal thermal gradients.
This setup could correspond to a linear chain of trapped ions as described in Refs. \cite{Muga1,Muga2}. 


We now discuss several aspects of TGHO chains:

(i) {\bf Absence of rectification with two affinities.}
If the beads at the thermostated segments are coupled to equal-temperature baths, 
$T_1=T_2$ and $T_4=T_5$ in Fig. \ref{Fig2},
then $\Delta J=0$ irrespective of structural asymmetry implemented via e.g. mass gradient, differing force constants or 
couplings to the baths. 

We emphasize that rectification does not develop in this single-affinity scenario 
even when the model is made more complex, e.g. by making the statistics of the baths quantum, 
including long-range (yet harmonic) interactions, or by allowing the baths to couple 
to all beads (with different strengths). 
This observation emerges from the analytic structure of the Landauer heat current expression.

(ii) {\bf Classical and quantum TGHO diodes.} 
As we showed in Eq. (\ref{eq:DeltaJ}), 
$\Delta J\neq 0 $ once the gradients are different, $(T_1-T_2) \neq (T_4-T_5)$,
unless a mirror symmetry is imposed with $M_{14} = M_{25}$. 
To break the symmetry between $M_{14}$ and $M_{25}$,
the thermostated regions should be made structurally asymmetric, i.e. $k_1\neq k_4$
 
(iii) {\bf  Purely-quantum TGHO diode.}
In the quantum limit, the temperatures in Eq. (\ref{eq:DeltaJ}) appear within
the Bose-Einstein distribution functions, included in the frequency integral. 
In this case, as long as at least three affinities are applied, e.g.
$T_1>T_2>T_4>T_5$, {\it and even when the gradients
are equal}, $(T_1-T_2)=(T_4-T_5)$, thermal rectification would show up 
(assuming structural asymmetry is included as required.)

(iv) {\bf Self consistent reservoir method.} The TGHO system is distinct from the self consistent reservoir (SCR) method, 
which was discussed in e.g. 
Refs. \cite{SC1,SC2,SC3,SC4a,SC4b,SC5} 
in the context of thermal rectification in quantum chains.
The role of the SCRs is to mimic anharmonicity. These fictitious thermal baths are attached to
 {\it interior} beads  while demanding zero net heat flow from the physical system to the SCRs. 
The temperature of the SCRs is dictated by this condition. 
In contrast, in the TGHO chain the thermostats are responsible for the power input and output 
from the system, and their temperature is freely assigned as independent boundary conditions.


\begin{figure} 
\includegraphics[width = .45\textwidth]{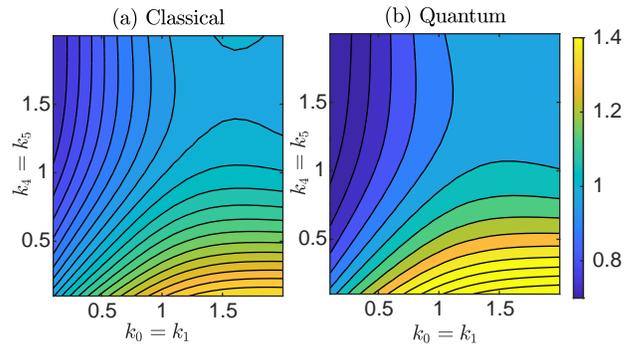}
\caption{
Contour plot of the rectification ratio in an $N=5$-particle harmonic chain with $N_H=N_C=2$.
(a) Classical case and (b) quantum calculation
with $T_1=1$, $T_2=0.5$, $T_4=0.2$, $T_5$=0.1 and $\gamma=1$; 
the central bead is not coupled to a thermostat.
We introduce different harmonic force constants at the thermostated regions, 
but use $k_2 = k_3 =1$ for the interior part, masses are set at $m=1$.}
\label{Fig3}
\end{figure}
\begin{figure} 
\includegraphics[width = .45\textwidth]{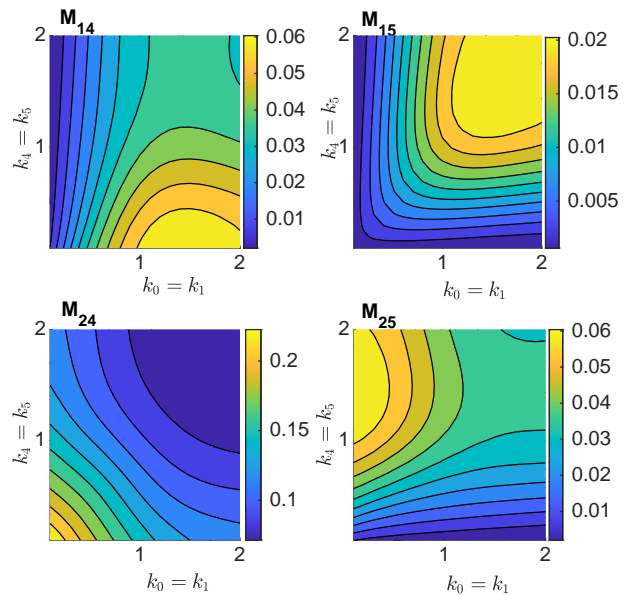} 
\caption{
The elements $M_{ij}$ for the classical model corresponding to simulations in Fig.~$\ref{Fig3}$.
Rectification arises due to the asymmetry $M_{14}\neq M_{25}$.
Parameters are the same as in Fig. \ref{Fig3}.}
\label{Fig4}
\end{figure}
\section{Simulations}

\subsection{Classical and Quantum TGHO diodes}

Rectification effect can be measured in different ways, 
with $\Delta J\neq 0$, defined in Eq. (\ref{eq:DeltaJ}), or based on a rectification ratio, $R\equiv |J/\tilde J|$.
We demonstrate the TGHO diode effect in Fig.
\ref{Fig3}, where we study the effect in the 5-bead system corresponding to Fig.~\ref{Fig2}.
We implement spatial asymmetry by using different force constants, $k_0 = k_1\neq k_4 = k_5$.
The current was calculated as the total input heat (\ref{eq:Jtotal}) (confirmed to be identical to the total heat dissipated to the cold baths) by numerically integrating Eqs. (\ref{eq:JC}) and (\ref{eq:JQ}) with a fine frequency grid up to a cutoff frequency larger than all other energy scales
For a discussion of the subtleties of the heat current definition see \cite{Naim}.

We show that both classical and quantum calculations can create the diode effect.
In Fig. \ref{Fig3}(a), the rectification ratio reaches up to $R\approx 1.4$ in both the classical and quantum cases.
 While the effect is not very large, it is in fact comparable to rectification ratios emerging due to an anharmonic potential, 
 as we discuss below in Fig.~(\ref{FKrect}).
In Fig. \ref{Fig3}(b) we display the behavior of the quantum TGHO diode, 
indicating on a somewhat stronger diode effect (bottom-right domain).

How can we tune the system to increase the rectification ratio? 
As can be seen from the analytic form of the heat current for a  5-bead chain,
there are four terms that play a role in the rectification ratio, 
$M_{14}$,  $M_{25}$, $M_{15}$ and $M_{24}$. 
These contributions are displayed in Fig. \ref{Fig4}. 
We conclude that at large asymmetry 
(bottom-right part), 
$M_{14}$  should dominate---once the gradients are made large. 
At this region, roughly $R \approx |(T_1-T_4)/(T_5-T_2)|$, which is $\approx 2$ in our parameters, 
close to the achieved maximal rectification ratio of 1.4.

Thus, a viable strategy to increase rectification is to
impose large structural asymmetry between the two ends, as well as apply 
significantly-unequal thermal gradients at the left and right side.
The large spatial asymmetry results in the the dominance of a single transport pathway. 
Furthermore, by imposing a large
 gradient at the left side, $\Delta T_H$, and a small gradient at the right side, 
$\Delta T_C$, with a small temperature drop on the central region 
(such that in the example used, $T_4\sim T_2$) the rectification ratio of the model scales as 
$R\propto|\Delta T_{H}/\Delta T_C|$. 
Below (Fig. \ref{lengthdep}) we further show that in long chains, the rectification effect is suppressed with $N_B=N_{H,C}$,
but it only weakly depend on $N_I$. We therefore suggest that 
$R\propto \frac{1}{N_B}\left|\frac{\Delta T_H}{\Delta T_C}\right|$. 

\begin{figure} [htb]
\includegraphics[width = .5\textwidth]{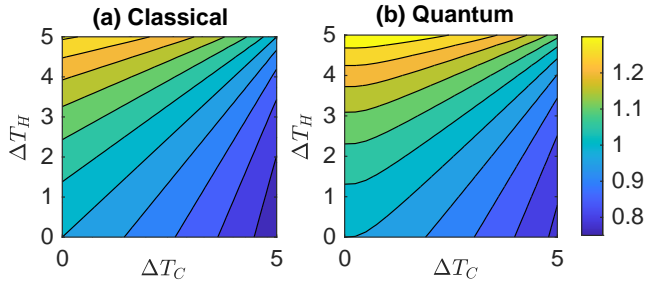} 
\caption{
Dependence of rectification on the temperature differences 
$\Delta T_H=T_1-T_2$ and $\Delta T_C=T_4-T_5$
in (a) classical and (b) quantum calculations.
Rectification is enhanced when one gradient is very large and the other small. 
Here, $T_1 = 10$, $T_2 = 10 - \Delta T_H$,
$T_4 = \Delta T_C$, $T_5 = 0$.
The force constants are $k_0 = k_1 = 2$, $k_2 = k_3 =1$, $k_4 = k_5 = .1$, $m=1$ and $\gamma = 1$.}
\label{tempgrad}
\end{figure}

\begin{figure} [htb]
\includegraphics[width = .33\textwidth]{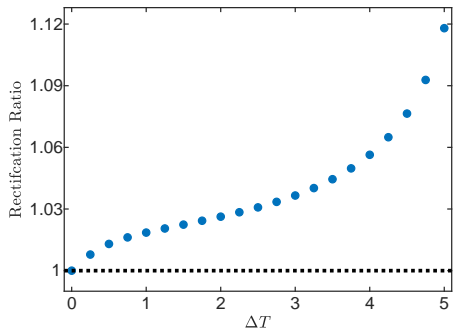} 
\caption{Purely-quantum TGHO diode operating when the thermal gradients at 
the two boundaries are equal, $\Delta T_H = \Delta T_C$; 
we display the diagonal of Fig.~(\ref{tempgrad}).
Parameters are $T_1 = 10$, $T_2 = 10 -\Delta T$, $T_4 = \Delta T$, $T_5 = 0$.}
\label{tempgradQ}
\end{figure}
\subsection{Purely-quantum TGHO diode}

The dependence of the classical and quantum TGHO diode effect on the local gradients is presented 
in Fig \ref{tempgrad}.
 As described above, the diode effect is enhanced when e.g. the 
left side experiences a large thermal gradient, while  temperatures at the right side are almost identical. 
The classical case cannot support the diode effect when the local gradients are equal, $\Delta T_H=\Delta T_C$.
In contrast, quantum statistics allows the diode behavior under equal gradients. 
This effect is illustrated in the behavior along the diagonal of Fig.~\ref{tempgrad}(b), presented for clarity
in Fig. \ref{tempgradQ}.


\subsection{Length dependence of the TGHO diode effect}

\begin{figure} 
\vspace{5mm}
\includegraphics[width = .45\textwidth]{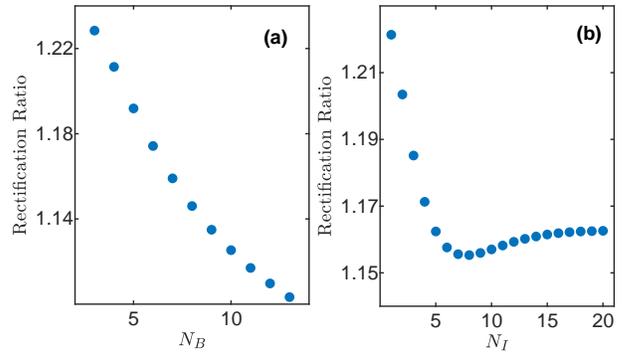} 
\caption{ 
Behavior of the rectification ratio with 
(a) $N_B$, number of thermostated sites, and (b) $N_I$, number of interior sites.
We set temperatures and gradients as $T_1=1$, $T_2=0.5$, $T_4=0.2$, $T_5=0.1$ thus
$\Delta T_H=1-0.5$ and $\Delta T_C=0.2-0.1$.
In panel (a), a linear gradient is assumed within each thermostated region. 
In panel (b), $N_H=N_C=2$. 
Other parameters are $\gamma = 1$, and force constants in the left (right) thermostated region at 1 (.1); other constants are set to 1. Simulations were performed using classical expressions.}
\label{lengthdep}
\end{figure}

Fig.~(\ref{lengthdep}) displays the behavior of the rectification ratio as the size of the 
system increases. 
In panel (a) we increase the number of thermostated sites $N_B$ while fixing the overall 
temperature differences $\Delta T_H$ and $\Delta T_C$, assuming a linear gradient in each region. 
We find that rectification decays as the number of thermostated sites increases. 
In contrast, the rectification ratio persists and saturates as we increase the number of 
sites in the interior region, $N_I$. This saturation is expected since in harmonic chains 
thermal transport is ballistic.
Thus, the impact of the central region on the the rectification effect should become 
independent of length, $N_I$ for long enough chains.

\subsection{Comparison to an anharmonic diode}

\begin{figure}
\vspace{3mm}
\includegraphics[width = .5\textwidth]{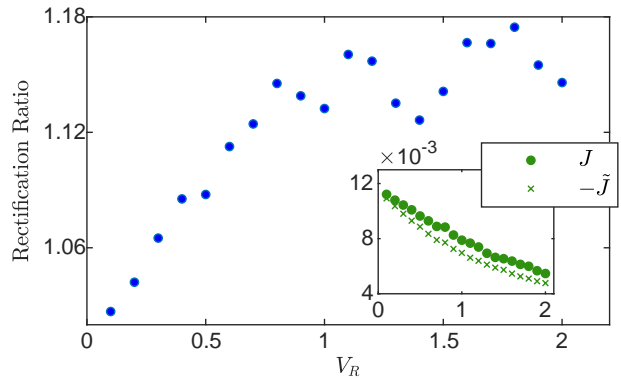} 
\caption{Rectification ratio in the Frenkel-Kontorova anharmonic chain.
The setup is analogous to Fig.~(\ref{Fig1})(a) with 
the leftmost and rightmost particles thermostated.
Rectification is achieved by adding different anharmonic onsite FK potentials 
to the left (first two beads) and right (last three beads) sides of the five-bead chain. 
Here, at the right side, $V_R = 1$ and $k=1$ while at the other half of the chain 
$V_L$ varies and $k=0.1$. 
Other parameters are $T_H = 1$, $T_C = .1$, $\gamma = 1$.
The inset presents the currents in the forward ($J$) and reversed ($\tilde J$) directions.}
\label{FKrect}
\end{figure}

To appreciate the magnitude of the rectification effect in the TGHO chain, we present in 
Fig.~(\ref{FKrect}) the diode behavior emerging when anharmonic interactions are explicitly added
to the chain.
We use the Frenkel-Kontorova (FK) potential that was used in many demonstrations of
nonlinear thermal devices, e.g., \cite{Li04,Li06,Li07}, adding onsite potentials to Eq. (\ref{eq:H}), 
 \bea 
 V(x) = V_{R/L}\cos\left(\frac{2\pi}{a}x\right).
 \eea 
Specifically, for the five-site chain, we encode asymmetry in the 
force constants and in the local potentials,
$V_L$ vs. $V_R$.
Unlike the harmonic case, which is analytically solvable, to treat anharmonic interactions we turn to numerical molecular dynamics simulations.
The Langevin equations of motion are integrated with the Br\"unger-Brooks-Karplus method; 
simulations were preformed by propagating the dynamics long enough to reach a steady state, then finding the heat current by averaging the local currents between adjacent beads. Here we compute heat current as the net power exchanged between central beads, 
$\langle J^C \rangle = \frac{k_2}{2}\langle(v_2+v_3)(x_3 - x_2 - a)\rangle$.
We then average over time and over 
realizations of the noise. Technical details were discussed in Ref. \cite{Naim}. 
Results are presented in Fig.~(\ref{FKrect}). 
Note that in the FK calculation, we resort to the standard modelling with a single thermal affinity,
$T_H$ at the left thermostat and $T_C$ at the right side.
Furthermore, only the leftmost (bead 1) and rightmost ($N$) beads are thermostated,

Comparing Fig.~(\ref{FKrect}) to e.g. Fig.~(\ref{Fig3}), we note that rectification in the anharmonic 
FK model is comparable to values received in the TGHO diode. 
Thus, while the rectification ratio demonstrated with the TGHO chain model is 
not impressive, is is similar to what one would achieve using similar parameters in the FK anharmonic chain, 
a central model for diodes examined in the literature. The FK model has been optimized to show large rectification ratio \cite{Li04}; similarly, it is interesting to explore means for enhancing the TGHO diode effect.

 
\section{Discussion and summary}
\label{Sec-summ}

We described a new type of a thermal diode, which is constructed in a purely-harmonic system when
attached to multiple thermostats, thus imposing at least two affinities. 
The TGHO diode operates when two conditions are met: The thermostated regions are 
 (i) structurally asymmetric with respect to each other 
and (ii) placed under unequal thermal gradients. 
We further proved the onset of a purely-quantum TGHO diode, which exists 
when the reservoirs (of different temperatures) are placed under equal gradients.
We analyzed the dependence of the TGHO diode effect on chain length and the applied
temperature gradient and further compared its performance to a 
diode model that was based on an anharmonic force field.

Recent studies used harmonic junctions with a single affinity 
(as in Figs. \ref{Fig1}(a)-(b))
to realize a diode effect \cite{Muga1,Muga2}; this was achieved by 
making parameters such as friction coefficients temperature dependent,
$\gamma_1(T)$, $\gamma_N(T)$.  We refer to such models as effective harmonic-oscillator diodes.
In this case, going back for simplicity to the classical limit, Eq. (\ref{eq:JC}),
the net heat current is given by
$J\propto (T_H-T_C) \gamma_1(T_H)\gamma_N(T_C)M_{1N}(T_H,T_C)$, where we extracted
the friction coefficients from the definitions of $M_{1N}$ in Eq. (\ref{eq:M}).
Assuming e.g. a linear dependence of friction coefficients
with the temperature of the attached bath, 
$\gamma_{1,N}(T_H)=\gamma_{1,N}+\lambda (T_H-T_C)$, 
and $\gamma_{1,N}(T_C)=\gamma_{1,N}-\lambda (T_H-T_C)$, with $\lambda$ as the slope,
one obtains a diode effect,
\bea 
\Delta J
\propto \lambda (T_H-T_C)^2 (\gamma_N-\gamma_1)M_{1N}(T_H,T_C),\eea
where for simplicity we assumed that the friction coefficients have a small effect on the the Green's function $G(\omega)$.
The diode effect $\Delta J\neq0$ relies on two conditions: 
(i) structural asymmetry in the form here of $\gamma_1\neq\gamma_N$,
and (ii) hidden-effective interactions $\lambda\neq0$, making parameters temperature dependent. 
Notably, 
this effective harmonic oscillator diode 
scales quadratically with the temperature difference,
$\Delta J\propto (T_H-T_C)^2$.
This quadratic scaling is the fingerprint of a hidden anharmonicity, illustrating
a nonlinear phenomena. 
In contrast, the TGHO diode is a linear effect, characterized by
the linear scaling of the net heat current with local 
temperature biases, $\Delta J\propto \Delta T$, see Eq. (\ref{eq:DeltaJ}).

Purely harmonic junctions connecting heat baths at two different temperatures cannot rectify heat. 
Our study shows that one may achieve a diode behavior in harmonic setups by using compound 
boundary conditions that enforce local thermalization on several sites.
Realizing a TGHO thermal diode with a large rectification ratio remains a challenge. 
Future work will be focused on testing the impact of long range interactions on the TGHO diode with
the goal to enhance its performance.

\begin{acknowledgments}
DS acknowledges the NSERC discovery grant and the Canada Chair Program. 
BKA gratefully acknowledges the start-up funding from IISER Pune, the MATRICS grant MTR/2020/000472 from SERB, Govt. of India,
and the hospitality of the Department of Chemistry at the University of Toronto.
BKA and DS thank the Shastri Indo-Canadian Institute for providing financial support for this research work in the form of a Shastri Institutional Collaborative Research Grant (SICRG).
\end{acknowledgments}

\section*{Appendix A: TGHO diode with asymmetric interparticle couplings}

\setcounter{equation}{0}
\renewcommand\theequation{A\arabic{equation}}

We show that rectification appears only when asymmetry is encoded such that the thermostated regions are distinct. 

For the five-site model with equal friction constants, $\boldsymbol{G^{-1}}(\omega)=$
\begin{widetext}
\bea
\begin{pmatrix} 
       -\omega^2 + i\gamma\omega + k_0 + k_1 & -k_1 \\
 -k_1 & -\omega^2 + i\gamma\omega + k_1 + k_2 & -k_2 \\
& -k_2 & -\omega^2 +                 k_2 + k_3 & -k_3 \\
&& -k_3 & -\omega^2 + i\gamma\omega + k_3 + k_4 & -k_4 \\
&&& -k_4 & -\omega^2 + i\gamma\omega + k_4 + k_5  \\
\nonumber\\
\end{pmatrix}
\eea
with zero elsewhere.
As discussed in the main text, in our setup, $N_H=N_C=2$, $N_I=1$; two beads are thermalized at the boundaries and a single bead at the centre is not directly coupled to heat baths. 
The diode effect for this system can be quantified by Eq. (\ref{eq:DeltaJ}), and it is controlled by the asymmetry between $M_{14}$ and $M_{25}$. We provide now explicit expressions for these terms, as defined in Eq. (\ref{eq:M}).

First,
 $G_{14} = \frac{\det(C_{14})}{\det(G^{-1})}$, where $C_{14}$ is the minor of $G^{-1}$, missing the row 1 and column 4,
\bea
C_{14} = 
\begin{pmatrix} 
-k_1 & -\omega^2 + i \gamma \omega + k_1 + k_2 & -k_2 & \\
& -k_2 & -\omega^2 + k_2 + k_3 & \\
& & -k_3  & -k_4 \\
& &  & -\omega^2 + i\gamma\omega+k_4 + k_5\end{pmatrix}
\eea
so
\bea
\det(C_{14})= -k_1k_2k_3(-\omega^2+i\gamma\omega + k_4 + k_5).
\eea
Similarly,
 \bea C_{25} = 
\begin{pmatrix} 
-\omega^2 + i\gamma\omega + k_0 + k_1 & -k_1 & &  \\ 
& -k_2 & -\omega^2 + k_2 + k_3 & -k_3  \\
& & -k_3 &-\omega^2 + i\gamma \omega + k_3 + k_4\\
& & &-k_4 \end{pmatrix}
\eea
so
\bea
\det(C_{25}) = -k_2k_3k_4(-\omega^2+i\gamma\omega + k_0 + k_1).
\eea
In order to obtain $\det(C_{14}) \neq \det(C_{25})$ we need to introduce an asymmetry, for example, setting $k_0 \neq k_5$ or $k_1 \neq k_4$. 
The parameters of the interior (unthermalized) region, $k_2$ and $k_3$,  play no role determining whether or not there will be rectification. Nevertheless, they can control the magnitude of the effect.
Explicitly,
 \bea M_{14} = \frac{\gamma^2}{\pi} \int_{-\infty}^{\infty} d\omega\omega^2\frac{\lvert k_1k_2k_3\left(-\omega^2 + i\gamma\omega + k_4 + k_5\right)\rvert^2}{\lvert\det \boldsymbol{G^{-1}}\rvert^2} 
 \eea
The denominator is a degree 20 polynomial of $\omega$. Its exact value
depends on all the system parameters. 
Other contributions to the current are given in terms of 
\begin{align}
\det(C_{15}) &= k_1k_2k_3k_4,\\
\nonumber \det(C_{24}) &= k_2k_3(-\omega^2+i\gamma\omega + k_0+k_1)(-\omega^2 + i\gamma\omega+k_4+k_5).
\end{align}
Longer chains have the analogous property that force constants between beads not connected to heat baths play no role in rectification: Asymmetry must appear between the sections directly thermalized by baths. 
%
More precisely, in an $N$-bead chain with $N_H=N_C=2$,  
so that beads 1, 2; $N-1$, and $N$ are connected to thermostats we get
\begin{align}
\det(C_{1(N)}) &= k_1k_{N-1}\left(\prod_{i=2}^{N-2}-k_i\right) \\ \nonumber
\det(C_{1(N-1)}) &= -k_1\left(\prod_{i=2}^{N-2}-k_i\right)(-\omega^2+i\gamma\omega+k_{N-1} + k_{N})\\ \nonumber
\det(C_{2(N)}) &= -k_{N-1}\left(\prod_{i=2}^{N-2}-k_i\right)(-\omega^2+i\gamma\omega+k_0 + k_1)\\ \nonumber
\det(C_{2(N-1)}) &= \left(\prod_{i=2}^{N-2}-k_i\right)(-\omega^2 + i\gamma\omega+k_0 + k_1)(-\omega^2+i\gamma\omega+k_{N-1} + k_N)
\end{align}
Rectification appears when $\det(C_{1(N-1)}) \neq \det(C_{2(N)})$, thus $k_1\neq k_{N-1}$ and/or $k_0\neq k_N$; asymmetry in the central region force constants, $k_2,\dots, k_{N-2}$ is not sufficient to enact rectification.
\end{widetext}

\section*{Appendix B: TGHO diode with asymmetric onsite potentials}
\setcounter{equation}{0}  
\renewcommand\theequation{B\arabic{equation}}
In this Appendix we include
asymmetry by introducing local trapping potentials with force constant $\tilde k$; the interparticle potentials are assumed identical.
For the five-particle chain with harmonic onsite potentials, the inverse Green's matrix has form 
\begin{widetext}
\bea
\boldsymbol{G^{-1}}(\omega) = 
\begin{pmatrix} 
     -\omega^2 + i\gamma\omega + 2k + \tilde k_1& -k \\
 -k & -\omega^2 + i\gamma\omega + 2k + \tilde k_2& -k \\
& -k & -\omega^2 +                 2k + \tilde k_3& -k \\
&& -k & -\omega^2 + i\gamma\omega + 2k + \tilde k_4& -k \\
&&& -k & -\omega^2 + i\gamma\omega + 2k + \tilde k_5 \\
\nonumber\\
\end{pmatrix}
\eea
We again set $N_H=N_C=2$ and $N_I=1$; two beads are thermalized at each boundary, while the single bead at the centre (particle 3) is not thermalized. The diode effect for this system can be quantified by Eq. (\ref{eq:DeltaJ}), and it is controlled by the asymmetry between $M_{14}$ and $M_{25}$. We provide now explicit expressions for these terms to analyze the required source of asymmetry.

The elements $\det(C_{ij})$ take the form
\begin{align}
\det(C_{15}) &= k^4,\\
\nonumber \det(C_{14})&= -k^3(-\omega^2+i\gamma\omega + 2k + \tilde k_5), \\
\nonumber \det(C_{25}) &= -k^3(-\omega^2+i\gamma\omega + 2k + \tilde k_1), \\
\nonumber \det(C_{24}) &= k^2(-\omega^2+i\gamma\omega + 2k + \tilde k_1)(-\omega^2 + i\gamma\omega + 2k+\tilde k_5).
\end{align}
In this case, rectification can show up  once $\tilde k_1 \neq \tilde k_5$ resulting in  $M_{14}\neq M_{25}$.
As in the case of asymmetric interparticle forces, this holds for chains of any size. 
For an $N$-bead chain with $N_H=N_C=2$,
\begin{align}
\det(C_{1(N)}) &= k^{N-1},\\
\nonumber \det(C_{1(N-1)})&= -k^{N-2}(-\omega^2+i\gamma\omega + 2k + \tilde k_N), \\
\nonumber \det(C_{2(N)}) &= -k^{N-2}(-\omega^2+i\gamma\omega + 2k + \tilde k_1), \\
\nonumber \det(C_{2(N-1)}) &= k^{N-3}(-\omega^2+i\gamma\omega + 2k + \tilde k_1)(-\omega^2 + i\gamma\omega + 2k + \tilde k_N).
\end{align}
\end{widetext}
Therefore, it is the asymmetry $\tilde k_1\neq \tilde k_N$ that is responsible for the diode effect.
Inspecting this form, we expect that the TGHO chain with an asymmetry in the interparticle force constants would support
larger rectification ratios than the case with pinning potentials.


\begin{thebibliography}{99}


\bibitem{RubtsovRev}
I. V. Rubtsov and A. L. Burin, 
Ballistic and diffusive vibrational energy transport in molecules, 
J. Chem. Phys. Perspective {\bf 150}, 020901 (2018).

\bibitem{ReddyRev}	
Thermal and thermoelectric properties of molecular junctions,
K. Wang, E. Meyhofer, and P. Reddy, 
Adv. Func. Mater. {\bf 30}, 1904534 (2020).

\bibitem{BijayRev}
D. Segal and B. K. Agarwalla,
Vibrational heat transport in molecular junctions,
Ann. Rev. Phys. Chem. {\bf 67}, 185 (2020).
%

\bibitem{DharRev}
A. Dhar, Heat transport in low dimensional systems,
Adv. Phys. {\bf 57}, 457 (2008).

\bibitem{LeitnerRev}
D. M. Leitner,
Quantum ergodicity and energy flow in molecules,
Adv.  Phys. {\bf 64}, 445 (2015).

\bibitem{FPU1}
E. Fermi, J. Pasta and S. Ulam,
Studies of nonlinear problems,
Los Alamos Scientific Laboratory report LA-1940 (1955).

\bibitem{FPU2}
 T. Dauxois,
Fermi, Pasta, Ulam, and a mysterious lady,
 Physics Today {\bf 61},  55 (2008). 
 
 \bibitem{Pauly}
J. C. Kl\"ockner and F. Pauly,
Variability of the thermal conductance of gold-alkane-gold single-molecule junctions studied using ab-initio and molecular dynamics approaches,
arXiv:1910.02443.


\bibitem{Braun}
M. D.  Losego, M. E. Grady, N. R. Sottos, D. G. Cahill, and P. V. Braun,
Effects of chemical bonding on heat transport across interfaces, 
Nat. Mater. {\bf 11}, 502 (2012).

\bibitem{Gotsmann}
 T. Meier, F. Menges, P. Nirmalraj, H. H\"olscher,  H. Riel, and B. Gotsmann 
 Length-dependent thermal transport along molecular chains, 
 Phys. Rev. Lett. {\bf 113}, 060801 (2014).
 
\bibitem{Malen}
S. Majumdar, J. A. Sierra-Suarez, S. N. Schiffres, W.-L. Ong, C. F. Higgs, A. J. H. McGaughey, and J. A. Malen,
Vibrational mismatch of metal leads controls thermal conductance of self-assembled monolayer junctions,
 Nano Lett. {\bf 15}, 2985 (2015).

\bibitem{Brendt}
N. Mosso, H. Sadeghi, A. Gemma, S. Sangtarash, U. Drechsler, C. Lambert, and B. Gotsmann,
Thermal transport through single-molecule junctions,
Nano Lett. {\bf 19}, 7614 (2019).
%

\bibitem{ReddyE}
L. Cui, S. Hur, Z. A. Akbar, J. C. Kl\"ockner, W. Jeong, F. Pauly, S. Y. Jang, P. Reddy, and E. Meyhofer,
Thermal conductance of single-molecule junctions,
Nature {\bf 572}, 628 (2019).

\bibitem{Lebowitz}
Z. Rieder, J. L. Lebowitz, and E. Lieb,
Properties of a Harmonic Crystal in a Stationary Nonequilibrium State,
J. Math. Phys. {\bf 8}, 1073 (1967).

\bibitem{Segal03}
D. Segal, A. Nitzan, and P. H\"anggi,
Thermal conductance through molecular wires,
J. Chem. Phys. {\bf 119}, 6840 (2003).

\bibitem{Rego98}
L. G. C. Rego and G. Kirczenow,
Quantized thermal conductance of dielectric quantum wires, 
Phys. Rev. Lett. {\bf 81}, 232 (1998).
%

\bibitem{commentMuga}
A thermal diode effect can be realized in ``effective" harmonic systems---once anharmonicity
is introduced at a mean field level, by making parameters temperature dependent
\cite{Pereirareq,Muga1,Muga2}.
In these type of models there is a cross-graining or mean-field
assumption making parameters temperature dependent. In Ref.
\cite{Muga2} for example  the friction coefficient was varied with temperature,
thus allowing the occurrence of a diode effect in an {\it effective} harmonic junction.
To eliminate confusion, we highlight that here all parameters are temperature independent, and that the
model is microscopically harmonic.

\bibitem{Pereirareq}
E. Pereira,
Requisite ingredients for thermal rectification,
Phys. Rev. E {\bf 96}, 012114 (2017).

\bibitem{Muga1}
M. A. Simon, S. Martinez-Garaot, M. Pons, and J. G. Muga,
Asymmetric heat transport in ion crystals,
Phys. Rev. E {\bf 100}, 032109 (2019).

\bibitem{Muga2}
M. A. Simón, A. Alaña, M. Pons, A. Ruiz-García, and J. G. Muga,
Heat rectification with a minimal model of two harmonic oscillators,
Phys. Rev. E {\bf 103}, 012134 (2021).


\bibitem{casati02}
M. Terraneo, M. Peyrard, and G. Casati, 
Controlling the energy flow in nonlinear lattices: A model for a thermal rectifier,
Phys. Rev. Lett. {\bf 88}, 094302 (2002).

\bibitem{Li04}
B. Li, L. Wang, and G. Casati,
Thermal Diode: Rectification of Heat Flux,
Phys. Rev. Lett. {\bf 93}, 184301 (2004).

\bibitem{BaowenRev}
N. Li, J. Ren, L. Wang, G. Zhang, P. H\"anggi, and B. Li,
Colloquium: Phononics: Manipulating heat flow with electronic analogs and beyond,
Rev.  Mod. Phys. {\bf 84}, 1045 (2012).

\bibitem{WongRev}
M. Y. Wong, C. Y. Tso, T. C. Ho, and H. H. Lee,
A review of state of the art thermal diodes and their potential applications
Int.l J. Heat and Mass Transfer {\bf 164}, 120607 (2021).


\bibitem{BenentiR}
G. Benenti,  G. Casati, C. Mejía-Monasterio, and M. Peyrard,
From Thermal Rectifiers to Thermoelectric Devices. In: Lepri S. (eds) Thermal Transport in Low Dimensions. Lecture Notes in Physics, vol 921. Springer, Cham. 2016.

\bibitem{Bijay17}
B. K. Agarwalla and D. Segal,
Energy current and its statistics in the nonequilibrium spin-boson model: Majorana fermion representation, New J. Phys. {\bf 19}, 043030 (2017).


\bibitem{SB}
D. Segal and A. Nitzan, 
Spin-boson thermal rectifier,
Phys. Rev. Lett. {\bf 94}, 034301 (2005).
%

\bibitem{PereiraRev}
E. Pereira,
Thermal rectification in classical and quantum systems: Searching for efficient thermal diodes,
Europhys. Lett. {\bf126}, 14001 (2019).

\bibitem{Leitner13}
D. M. Leitner,
Thermal boundary conductance and thermal rectification in molecules
J. Phys. Chem. B {\bf 117}, 12820 (2013).

\bibitem{Leitner19}
K. M. Reid, H. D. Pandey, and D. M. Leitner,
Elastic and inelastic contributions to thermal transport between chemical groups and thermal rectification in molecules,
J. Phys. Chem. C {\bf 123}, 6526 (2019).


\bibitem{CasatiEPL}
S. Chen, E. Pereira, and G. Casati,
Ingredients for an efficient thermal diode,
EPL  {\bf 111},  30004 (2015).

\bibitem{Chen}
S. Chen, D. Donadio, G. Benenti, and G. Casati,
Efficient thermal diode with ballistic spacer
Phys. Rev. E {\bf 97}, 030101 (2018).


\bibitem{Alexander}
T. Alexander,
High-heat-flux rectification due to a localized thermal diode,
Phys. Rev. E {\bf 101}, 62122 (2020).

\bibitem{You}
S. You, D. Xiong, and J. Wang,
Thermal rectification in the thermodynamic limit,
Phys. Rev. E {\bf 101}, 012125 (2020).


\bibitem{Wu09}
S.-A. Wu and D. Segal,
Sufficient conditions for thermal rectification in hybrid quantum structures,
Phys. Rev. Lett. {\bf 102}, 095503 (2009).


\bibitem{Claire09}
L.-A. Wu, C. X. Yu, and D. Segal,
Nonlinear quantum heat transfer in hybrid structures: Sufficient conditions for thermal rectification,
Phys. Rev. E {\bf 80}, 041103 (2009). 

\bibitem{XXP}
S. H. S. Silva, G. T. Landi, R. C. Drumond, and E. Pereira,
Heat rectification on the XX chain,
Phys. Rev. E {\bf 102}, 062146 (202).


\bibitem{Vinitha}
V. Balachandran, G. Benenti, E. Pereira, G. Casati, and D. Poletti,
Heat current rectification in segmented XXZ chains,
Phys. Rev. E {\bf 99}, 032136 (2019).


\bibitem{Goold}
V. Balachandran, S. R. Clark, J. Goold, and D. Poletti,
Energy current rectification and mobility edges,
Phys. Rev. Lett. {\bf 123},  020603 (2019).

\bibitem{Pekola}
J. Senior, A. Gubaydullin, B. Karimi, J. T. Peltonen, J. Ankerhold, and J. P. Pekola,
Heat rectification via a superconducting artificial atom,
Comm. Phys. {\bf 3}, 40 (2020).
%

\bibitem{Haack}
A. Iorio, E. Strambini, G. Haack, M. Campisi, and F. Giazotto,
Photonic heat rectification in a coupled qubits system,
arXiv:2101.11936.

\bibitem{Dlott1}
B. C. Pein, Y. Sun, and D. D. Dlott,
Unidirectional vibrational energy flow in nitrobenzene,
J . Phys. Chem. A {\bf 117}, 6066 (2013).

    
\bibitem{Dlott2}
B. C. Pein, Y. Sun, and D. D. Dlott,
Controlling vibrational energy flow in liquid Alkylbenzenes,
J. Phys. Chem. B {\bf 117}, 10898  (2013).

\bibitem{CNTE}
Nanoscale solid-state thermal rectifiers were realized e.g. in 
Ref. \cite{Zettl} based on mass-graded carbon and boron nitride nanotubes.

\bibitem{Zettl}
 C. W. Chang, D. Okawa, A. Majumdar, and A. Zettl,
Solid-state thermal rectifier, 
Science  {\bf 314}, 1121 (2006).

\bibitem{Haffner}
M. Ramm, T. Pruttivarasin, and H. H\"affner,
Energy transport in trapped ion chains,
New J. Phys. {\bf 16}, 063062 (2016).


\bibitem{Inon1}
I. Sharony, R. Chen, and A. Nitzan,
Stochastic simulation of nonequilibrium heat conduction in extended molecule junctions,
J. Chem. Phys. {\bf 153}, 144113 (2020).

\bibitem{SC1}
F. Bonetto, J. L. Lebowitz, and J. Lukkarinen, 
Fourier's Law for a harmonic crystal with self-consistent stochastic reservoirs,
J. Stat. Phys. {\bf 116}, 783 (2004).

\bibitem{SC2}
E. Pereira and H. C. F. Lemos, 
Symmetry properties of heat conduction in inhomogeneous materials,
Phys. Rev. E {\bf 78}, 031108 (2008).

\bibitem{SC3}
D. Segal,
Absence of thermal rectification in asymmetric harmonic chains with self consistent reservoirs: An exact analysis,
Phys. Rev. E {\bf 79}, 012103 (2009).

\bibitem{SC4a}	
M. Bandyopadhyay and D. Segal,
Quantum heat transfer in harmonic chains with self consistent reservoirs: Exact numerical simulations,
Phys. Rev. E {\bf 84}, 011151 (2011).
	
\bibitem{SC4b}
E. Pereira, H. C. F. Lemos, and R. R. Ávila,
Ingredients of thermal rectification: The case of classical and quantum self-consistent harmonic chains of oscillators,
Phys. Rev. E {\bf 84}, 061135 (2011).

\bibitem{SC5}
R. Moghaddasi Fereidani and D. Segal,
Phononic heat transport in molecular junctions: quantum effects and vibrational mismatch,
J. Chem. Phys. {\bf 150}, 024105 (2019).

\bibitem{Naim}
N. Kalantar, B. K. Agarwalla and D, Segal,
On the definitions and simulations of vibrational heat transport in nanojunctions,
J. Chem. Phys. {\bf 153}, 174101 (2020).

\bibitem{Li06}
B. Li, L. Wang, and G. Casati,
Negative differential thermal resistance and thermal transistor,
App. Phys. Lett. {\bf 88}, 143501 (2006).

\bibitem{Li07}
L. Wang and B. Li,
Thermal logic gates: computation with phonons,
Phys. Rev. Lett. {\bf 99}, 177208 (2007).







\end{thebibliography}
\end{document}